\documentclass{article}

\usepackage{arxiv}

\usepackage[utf8]{inputenc} 
\usepackage[T1]{fontenc}    
\usepackage{hyperref}       
\usepackage{url}            
\usepackage{booktabs}       
\usepackage{amsfonts}       
\usepackage{nicefrac}       
\usepackage{microtype}      
\usepackage{lipsum}		
\usepackage{graphicx}
\usepackage{natbib}
\usepackage{doi}
\usepackage{comment}
\usepackage{array}
\usepackage{multirow}
\usepackage{makecell}
\usepackage{tikz}
\usepackage{textcomp}
\usepackage{hyperref}
\usepackage{lipsum}

\newcommand\copyrighttext{%
  \footnotesize \textcopyright 2022 IEEE. Personal use of this material is permitted.
  Permission from IEEE must be obtained for all other uses, in any current or future media, including reprinting/republishing this material for advertising or promotional purposes, creating new collective works, for resale or redistribution to servers or
  lists, or reuse of any copyrighted component of this work in other works.
}
\newcommand\copyrightnotice{%
\begin{tikzpicture}[remember picture,overlay]
\node[anchor=south,yshift=10pt] at (current page.south) {\fbox{\parbox{\dimexpr\textwidth-\fboxsep-\fboxrule\relax}{\copyrighttext}}};
\end{tikzpicture}%
}

\title{A Qualitative Evaluation of Service Mesh-based \\Traffic Management for Mobile Edge Cloud
\thanks{This work was partially supported by the Wallenberg AI, Autonomous Systems and Software Program (WASP) funded by the Knut and Alice Wallenberg Foundation.}}

\author{Aleksandra Obeso Duque\\
	Department of Computing Science, Umeå University\\
	Cloud Systems and Platforms, Ericsson Research\\
	Stockholm, Sweden\\
	\texttt{alekodu@cs.umu.se} \\
	\And
	Cristian Klein \\
	Department of Computing Science\\
	Umeå University\\
	Umeå, Sweden\\
	\texttt{cklein@cs.umu.se} \\
	\AND
	Jinhua Feng\\
	Cloud Systems and Platforms\\
	Ericsson Research\\
	Stockholm, Sweden\\
    \texttt{jim.feng@ericsson.com} \\
	\And
	Xuejun Cai\\
	Cloud Systems and Platforms\\
	Ericsson Research\\
	Stockholm, Sweden\\
	\texttt{xuejun.cai@ericsson.com}\\
	\And
	Björn Skubic\\
	Cloud Systems and Platforms\\
	Ericsson Research\\
	Stockholm, Sweden\\
	\texttt{bjorn.skubic@ericsson.com}\\
	\And
	Erik Elmroth\\
	Department of Computing Science\\
	Umeå University\\
	Umeå, Sweden\\
	\texttt{elmroth@cs.umu.se}\\
}

\renewcommand{\headeright}{CCGRID2022}
\renewcommand{\undertitle}{CCGRID2022}
\renewcommand{\shorttitle}{Service Mesh-based Traffic Management for Mobile Edge Cloud}

\hypersetup{
pdftitle={A template for the arxiv style},
pdfsubject={q-bio.NC, q-bio.QM},
pdfauthor={David S.~Hippocampus, Elias D.~Striatum},
pdfkeywords={First keyword, Second keyword, More},
}

\begin{document}
\maketitle
\copyrightnotice

\begin{abstract}
Service mesh is getting widely adopted as the cloud-native mechanism for traffic management in microservice-based applications, in particular for generic IT workloads hosted in more centralized cloud environments. Performance-demanding applications continue to drive the decentralization of modern application execution environments, as in the case of mobile edge cloud.

This paper presents a systematic and qualitative analysis of state-of-the-art service mesh to evaluate how suitable its design is for addressing the traffic management needs of performance-demanding application workloads hosted in a mobile edge cloud environment. With this analysis, we argue that today's dependability-centric service mesh design fails at addressing the needs of the different types of emerging mobile edge cloud workloads and motivate further research in the directions of performance-efficient architectures, stronger QoS guarantees and higher complexity abstractions of cloud-native traffic management frameworks.
\end{abstract}


\section{Introduction}
\label{intro}

The advanced digitalization of industry, enterprise, and society drives increasingly strict and stringent application performance requirements compared to what is offered by Centralized Cloud (CC) execution environments. The always demanding application performance requirements motivate the distribution and decentralization of computational resources, as in the case of Edge Cloud (EC).

\begin{figure}[htbp]
	\centering
	\includegraphics[width=0.5\linewidth]{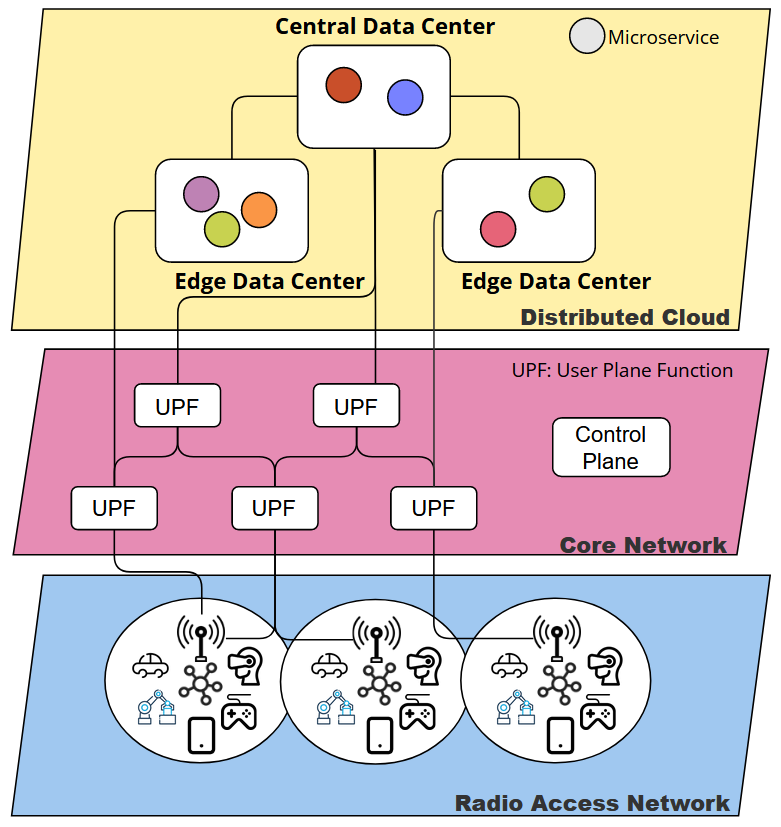}
	\caption{Mobile Edge Cloud architecture. The Mobile Network (MN) is composed by the Radio Access Network (RAN) and the Core Network (CN). The CN is, at the same time, composed by a Control Plane (CP) and a User Plane Function (UPF). On the other hand, the application execution environment is distributed across Centralized Cloud (CC) and Edge Cloud (EC) clusters.}
	\label{fig:mec}
\end{figure}

EC allows applications to perform better and, at the same time, reduces network congestion compared to more CC environments. Mobile Edge Cloud (MEC), a.k.a. Multi-access Edge Computing, is a type of EC where the computing nodes are placed geographically close to Mobile Network (MN) nodes, and thus closer to mobile end-users (see Figure~\ref{fig:mec}). Hence, MEC introduces new opportunities to optimize application performance at the cost of an increasing degree of complexity.

\begin{figure}[htbp]
	\centering
	\includegraphics[width=0.7\linewidth]{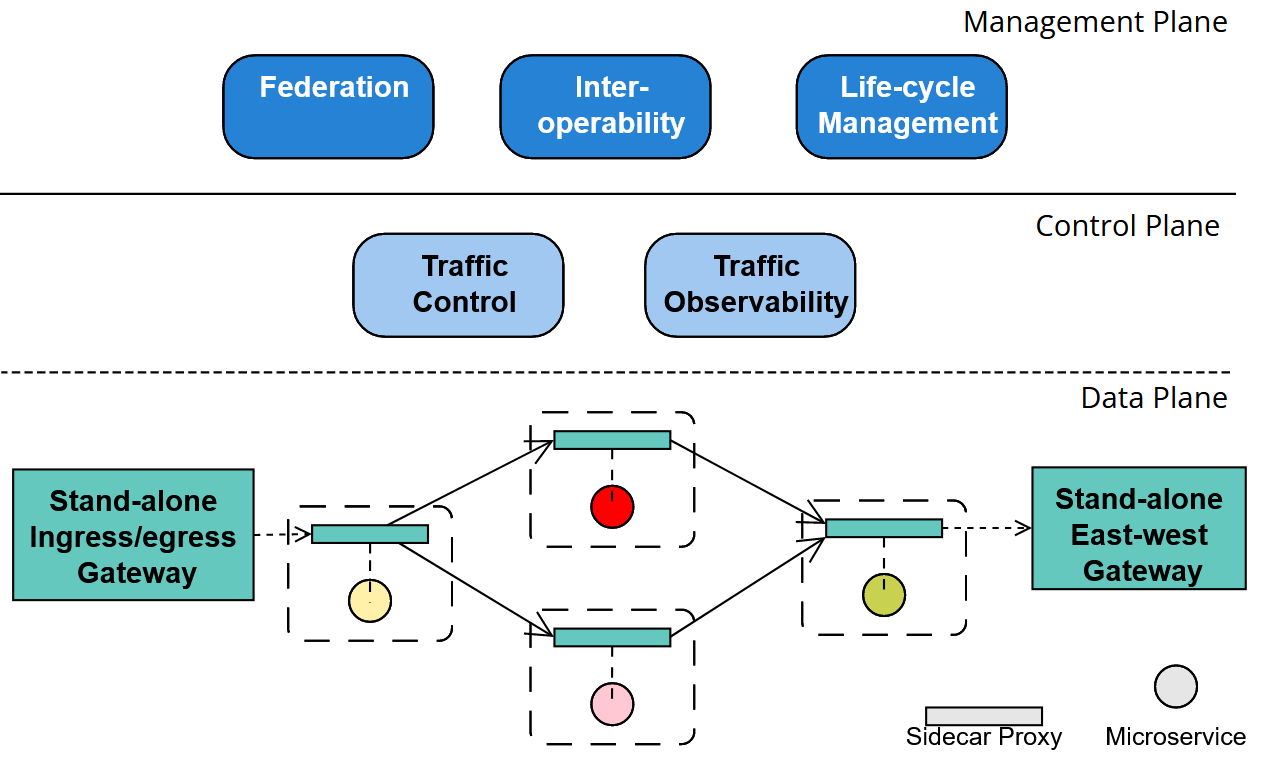}
	\caption{General Service Mesh functionality and architecture. The Service Mesh (SM) is generally split into a Management Plane (MP), a Control Plane (CP) and a Data Plane (DP).}
	\label{fig:general-architecture-service-mesh}
\end{figure}

Managing service-to-service communication in a microservice-based architecture is a challenging task. The microservice-based architecture has several advantages, but it also brings complexities related to the development and configuration of disaggregated application components, especially when dealing with a high number of very distributed microservices \cite{ccgrid-routing}. Application-level Service Mesh (SM) is being proposed as a cloud-native approach for traffic management which is getting widely adopted as a de-facto standard. SM supports configurable traffic control, consistent traffic observability, among others. A generic SM is realized as a dedicated infrastructure layer, unifying and centralizing the management and operation of microservice communication (see Figure~\ref{fig:general-architecture-service-mesh}).

However, the design of SM technology is currently oriented towards the needs of generic IT workloads, such as web-based applications, intended to be hosted in CC execution environments. These applications have relatively low performance demands when compared with emerging applications such as the ones targeting automotive, industrial control, Augmented Reality/Virtual Reality (AR/VR) and Internet of Things (IoT) use-cases; meanwhile the latter ones have strong demands in terms of latency, bandwidth and/or reliability. To know if SM can adequately be used for traffic management in MEC, there is a need to analyze in detail their current design. 

The research question addressed by this paper is: \textit{How suitable is the service mesh design for traffic management of performance-demanding applications hosted in a mobile edge cloud?} To answer this question, we perform a systematic and qualitative evaluation of state-of-the-art (SoTA) SM design for addressing the traffic management needs of performance-demanding applications running on top of MEC. We do not aim to base our analysis on a comparison across existing SM implementations, but rather on overall SoTA design features of SM.

We start by reviewing related work in this area (see Section~\ref{relatedWork}). Then, we define a set of evaluation criteria based on the characteristics and requirements from MEC and its application workloads (see Section~\ref{evaluationCriteria}). Such evaluation criteria are contrasted with the design drivers, functional and non-functional characteristics of SoTA SM (see Section~\ref{SM}). As a result of this evaluation, we identify SM strengths, limitations and trade-offs that motivate further research in the area (see Section~\ref{conceptualEvaluation}).

Our main contributions are two-folded: i. identification of strengths, limitations and tradeoffs associated with SM for MEC, and ii. identification of research challenges and opportunities related to the MEC-specialized SM we envision. With our analysis, we argue why SM fails at addressing the needs of emerging application workloads and foster further research in the areas of performant architecture, differentiated Quality of Service (QoS) assurance, and autonomy of SM. These contributions may represent critical input to standardization efforts related to cloud-native traffic management frameworks not only in the open-source community, i.e., the Service Mesh Interface (SMI)\footnote{https://layer5.io/projects/service-mesh-interface-conformance, The service mesh interface, accessed 2021-10-04}, but also the ones related to 3GPP standards for next-generation mobile networks, in particular, enhanced application architecture for enabling edge applications \cite{3gpp}.

\section{Related Work}
\label{relatedWork}

Around 2016, one of the first, well-known and well-adopted open-source SM implementations, Linkerd v1, was proposed. Istio and Linkerd v2 started to emerge in 2017 and 2018, respectively. Interesting to note that, at the time of writing, there are more than 20 different implementations of SM\footnote{https://layer5.io/service-mesh-landscape, The SM landscape: Comparison of service mesh strengths, accessed 2021-10-04}. Thus, the SMI initiative is aiming at standardizing the core functionalities of SM. In terms of SM management and operation, tools such as Gloo Mesh and Meshery are starting to appear.

However, most of the SM concepts and implementations are led by the open-source community. Interest from the research community started to emerge in 2017. This section presents related research work in terms of enhanced performant architecture, telco applicability and self-adaptability associated with SM, in relation to our insights discussed in Section \ref{conceptualEvaluation}.

\subsection{Performant Service Mesh Architecture}

In terms of performance enhancement, one proxy-less approach has been proposed. Subhraveti, D., et al. propose AppSwitch, a transport layer network element for modern application architectures \cite{appswitchid}. AppSwitch provides common network connectivity functions such as consistent application identity, application firewall and load balancing logic \cite{appswitchlb} without adding extra cost or complexity. It is implemented as a system call trap/generation function that propagates application location information via a gossip protocol.

\subsection{Service Mesh for Telco Domain}

Initial efforts are also identified in terms of SM for telco use-cases. For example, Dobaj, J., et al. identify and summarize relevant challenges regarding dependable network design for mission-critical systems, i.e., the need of ensuring dependable system and service connectivity, highly dynamic and flexible connectivity, and holistic and system-wide design approaches to support system interoperability \cite{10.1145/3361149.3361174}. The authors evaluate three networking patterns: the physical mesh pattern, the logic mesh pattern and the SM pattern for microservice-based applications.

Furthermore, Li, W., et al. explore the challenges, SoTA and future research opportunities of SM \cite{8705911}. This survey paper revises SM concepts and implementations highlighting opportunities for future research in the area, in particular, the need for further exploration of SM applicability into edge computing environments.

On the other hand, Antichi, G. and Rétvári, G. explore the idea of a full-stack Software-defined Networking (SDN) framework in relation to SM \cite{10.1145/3373360.3380834}. They highlight the need for substantial research in the area, in terms of performant and adaptable architectures and identify high-level SM limitations such as lack of support for carrier-grade performance, QoS assurance and compliance with telco standards and protocols.


\subsection{Self-adaptability of Service Mesh}

In terms of generic self-adaptation support in SM, initial efforts are seen in the research community. For example, Mendon\c{c}a, N., et al. study the lack of adoption of existing self-adaptation frameworks in the industry community \cite{10.1145/3241403.3241423}. The paper presents a survey and evaluation of existing solutions in terms of generality vs. reusability and identify multiple self-adaptation patterns, i.e., system-level, infrastructure-level and cross-layer patterns. Furthermore, the authors identify SM as a promising uniform architectural style to handle life cycle of self-adaptive distributed applications. 

Saleh Sedghpour, M. R., et al. study and enhance the circuit breaking capability of existing SM \cite{10.1145/3447851.3458740}. They argue that existing SM lacks adaptable circuit breaking actuation logic. To tackle this problem, they propose an adaptive controller that avoids overload, mitigates failures, and keeps tail response time below an specified threshold, while keeping throughput at a maximum. To do so, this controller dynamically adjusts circuit breaking queue length thresholds. The proposed adaptive controller can be easily configured to optimize the tradeoff between response time and throughput in a customized way.

Kosińska, J. and Zieliński, K. evaluate how important autonomous performance management is for addressing the highly dynamic requirements characteristic of a cloud-native execution environment \cite{kz20}. The paper addresses the design and evaluation of a technology-agnostic framework for autonomic management of cloud-native applications, which allows dynamic and on-the-fly reconfiguration. They also compare the proposed framework vs. SM from a design standpoint and identify SM limitations in terms of observability, communication patterns and automated control. 

Furthermore, Larsson, L., et al. enhance the resiliency capability of SM-based architectures by adding an adaptive and application-agnostic inter-service response caching mechanism \cite{9492576}. Their proposal is implemented as gRPC interceptors taking care of estimating response longevity and caching. With this enhanced mechanism, they  achieve 40\% traffic reduction and successful caching for about 80\% responses.

On the other hand, Mendon\c{c}a, N. C., et al. investigate cloud-native self-adaptive SM frameworks by revisiting the history of architectural connectors. These connectors are classified into five different generations, where the fifth one corresponds to the fully-fledged service communication platform, i.e., SM \cite{10.1145/3474624.3477072}. The authors argue that none of these generations provide direct support for changing connector's behavior at runtime. They envision software connectors supporting self-adaptation capabilities and allowing operators to customize the logic based on application constraints. They are, at the time of writing, building a prototype of a self-adaptive SM called KubowMesh.

In contrast to the prior knowledge presented in this section, our work focuses on a more thorough evaluation of today's SM design for the specific context of MEC and its different types of workloads, based on its design drivers, and functional and non-functional characteristics. As a result of our work, we identify additional challenges and devise future research directions in this area; especially, for differentiated QoS support as a new feature of a MEC-specialized SM.

\section{Evaluation Criteria from MEC and its Workloads}
\label{evaluationCriteria}

In the following subsections, we identify and analyze characteristics and requirements from MEC infrastructure and its different types of application workloads, as the criteria we use in Section~\ref{conceptualEvaluation} for evaluating SoTA SM. The identified characteristics and requirements are used to build profile models for MEC workloads and infrastructure, which are respectively depicted in figures \ref{fig:req-MEC-workloads} and \ref{fig:reqMEC}. These diagrams are inspired by SysML requirement diagrams\footnote{https://sysml.org/sysml-faq/what-is-requirement-diagram.html, What is a SysML Requirement diagram?, accessed 2021-10-12}, a general-purpose architecture modeling language for software requirement engineering.

To build the model depicted in Figure~\ref{fig:req-MEC-workloads}, we categorize application workloads into three different types of profiles based on their performance requirements: Mission-critical Applications, Bandwidth-demanding Applications and Latency-sensitive Applications. For each of these profiles we identify performance constraints and their associated requirements. Note that we mainly focus on third-party applications that would most likely be provided by non-telco enterprises, i.e., we do not focus on the considerations from cloud-native network functions as such. To build the MEC Infrastructure profile depicted in Figure~\ref{fig:reqMEC}, we identify MEC design constraints, design problems and their implicated requirements.

\subsection{Mission-critical Applications}
Mission-critical applications perform essential operations for their users and may have extremely high-cost or irreparable losses in case of failure, therefore they have high demands in terms of reliability. In general, users expect to have a high degree of dependability on such applications; thus, there is very low or zero tolerance against application errors or downtime. To deal with such requirements, the applications and the underlying system need to support very high availability, reliability and resiliency/survivability which may be, e.g., achieved based on redundancy and replication strategies (see Figure~\ref{fig:req-MEC-workloads}). Examples of such applications are remote operation and control of transport systems.

\subsection{Bandwidth-demanding Applications}
Bandwidth-demanding applications require the network to transfer high volume of data at very high rates. To achieve those demands, network links need to have enough capacity and advanced mechanisms for traffic management especially when such links are shared with other application loads. An important and complex aspect of this type of applications is the dynamicity of the bandwidth consumption imposed by variable traffic loads and even more, the case of bursty load in which the bandwidth demand is instantaneously spiked up leading to resource starvation (see Figure~\ref{fig:req-MEC-workloads}). Examples of such applications are data collection and processing from IoT sensors.

\subsection{Latency-sensitive Applications}
\begin{figure}[htbp]
	\centering
	\includegraphics[width=0.7\linewidth]{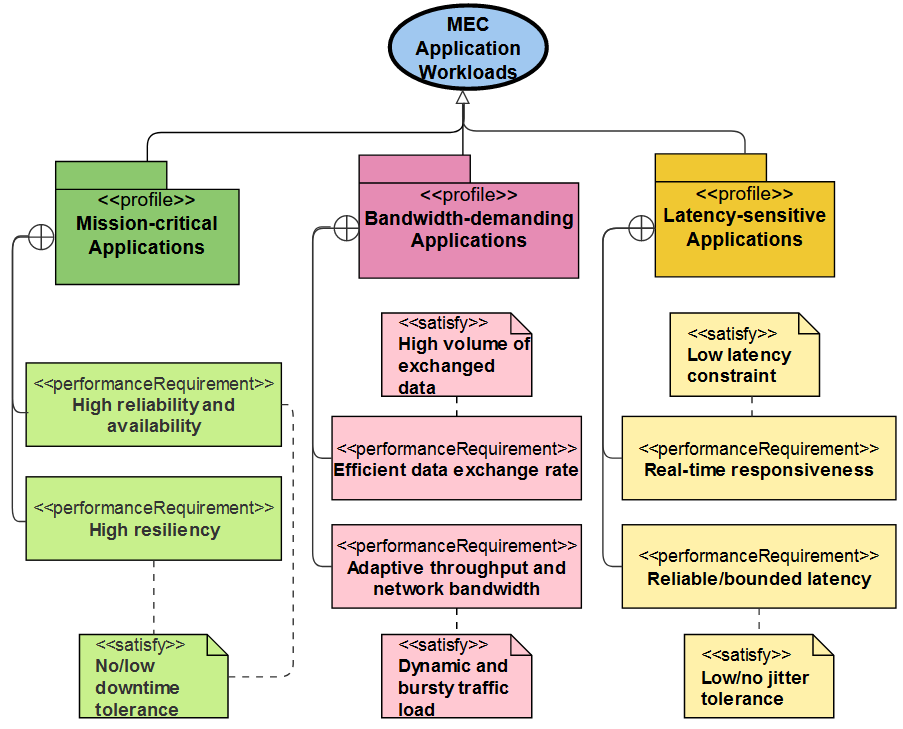}
	\caption{Evaluation Criteria: Characteristics and requirements of different types of performance-demanding applications.}
	\label{fig:req-MEC-workloads}
\end{figure}

Latency-sensitive applications require very low delays since they have very low time frames in which responses are expected. In general, they require high application responsiveness. The end-to-end time constraint can be defined in units of milliseconds or even hundreds of microseconds. To deliver such high responsiveness, these applications need to be supported by real-time software and networks. A key aspect to be supported is reliable/bounded latency; in such case it is important to reduce or eliminate jitter (see Figure~\ref{fig:req-MEC-workloads}). Examples of such applications are cloud gaming, AR/VR and industrial remote control.

\subsection{Mobile Edge Cloud Infrastructure}

Highly distributed cloud offers several advantages in terms of e.g., augmented overall computational capacity for large-scale application deployments and geographical redundancy for diverse failover scenarios. When compared to CC, the individual EC sites can also be considered more resource-constrained. Furthermore, EC presents higher resource and performance heterogeneity.

MEC nodes are located in proximity with access network nodes, and thus end-users; this translates into lower transport delays and less network congestion. Distributed cloud allows applications to perform better. MEC can host performance-demanding workloads from third-party application developers or content providers, but can also host Cloud-native Network Functions (CNF) from network providers/operators. To provide optimized QoS, the MN implements differentiated QoS assurance mechanisms based on, e.g., traffic flows, network slicing and Service Function Chaining (SFC) for optimized traffic steering.

However, to provide enhanced end-to-end performance guarantees, it is not enough to merely deploy EC data centers geographically closer to the MN's points of presence. It is also required to have tight integration, awareness, alignment and coordination between both technology domains, the MN and the EC. Performance-demanding applications impose the need for more holistic traffic management approaches, with federation and interoperability considerations.

\begin{figure}[htbp]
	\centering
	\includegraphics[width=0.7\linewidth]{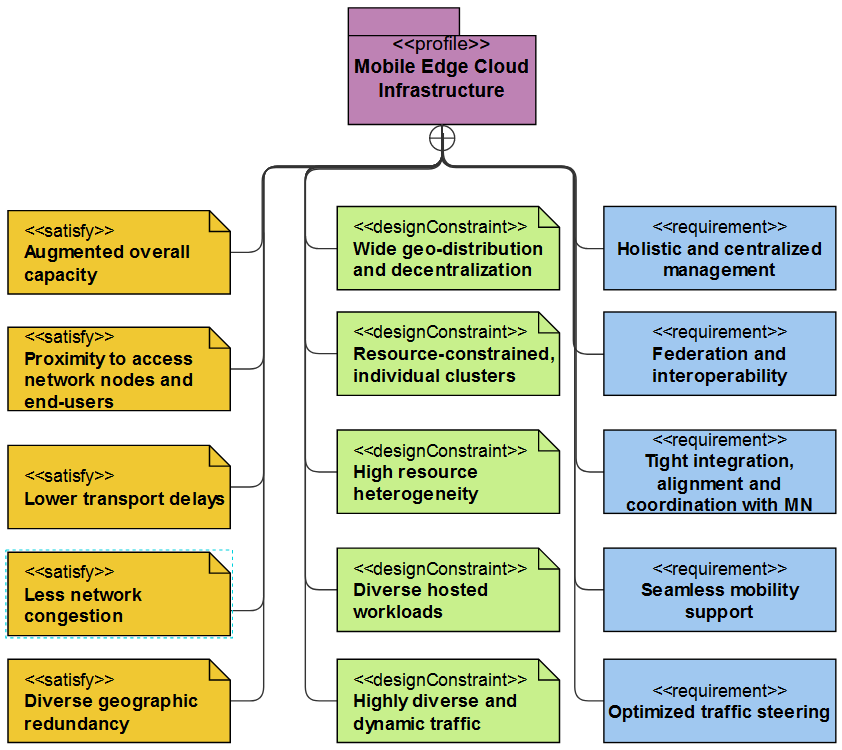}
	\caption{Evaluation Criteria: Characteristics and requirements of Mobile Edge Cloud infrastructure.}
	\label{fig:reqMEC}
\end{figure}

On the other hand, MN traffic can be considered highly diverse and highly dynamic, not only in terms of the traffic load itself but also due to the need of providing mobility support. Mobile end-users expect to have seamless, uninterrupted, non-degraded Quality of Experience (QoE), even upon mobility events. Figure~\ref{fig:reqMEC} summarizes MEC's characteristics and requirements.

\section{Service Mesh Characteristics and Design Drivers}
\label{SM}

In this section, we identify functional and non-functional characteristics of SoTA SM and its associated design drivers which we use to build the block diagrams depicted in figures \ref{fig:SMFunct} and \ref{fig:SMNonFunct}. In Section~\ref{conceptualEvaluation}, these design characteristics are evaluated based on the criteria identified in Section~\ref{evaluationCriteria}.

To build the functional diagram depicted in Figure~\ref{fig:SMFunct}, we categorize collected design features into conditions/options of control, actuation mechanisms and observability mechanisms. In the case of the diagram depicted in Figure~\ref{fig:SMNonFunct}, we identify non-functional characteristics and for each of them we specify the mean or mechanism on which such characteristic relies on. We also identify design drivers and their associated use-cases.

Note that we do not perform a comparison across existing SM implementations; we rather collect, to the best of our knowledge, the full set of features of SoTA SM design. To a great extent, our analysis is based on features from Istio since it is one of the most well-adopted, mature and feature-full implementations of SM, in particular due to its support for multi-cluster, multi-network deployment models needed in the MEC context. However, we also consider features from other well-adopted SM implementations such as Linkerd, Consul and Kuma plus additional features that are not currently supported by such implementations, i.e., the ones proposed by the research community. 

\subsection{Functional Characteristics}

\begin{figure}[htbp]
	\centering
	\includegraphics[width=0.7\linewidth]{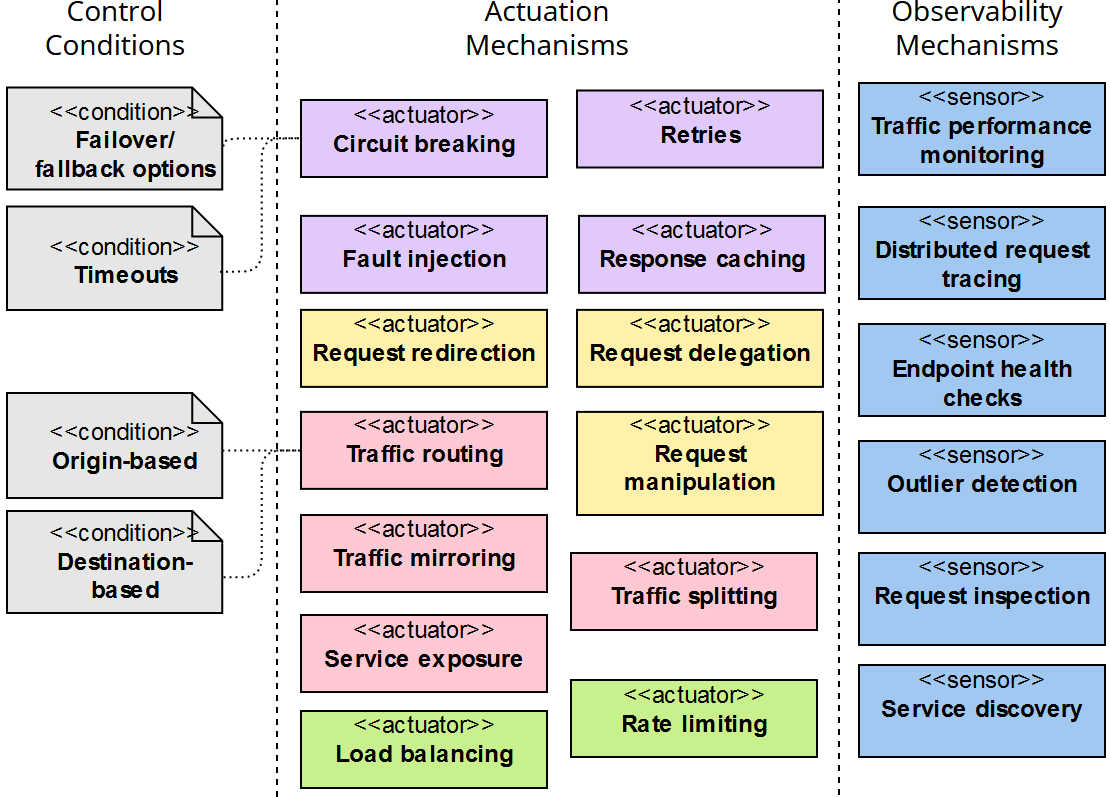}
	\caption{Functional characteristics of SoTA Service Mesh. Actuators in yellow and red are basic traffic control mechanisms, the ones in purple are related to reliability concerns, and the ones in green are related to bandwidth concerns.}
	\label{fig:SMFunct}
\end{figure}

In terms of traffic management, SM provides a set of functional features related to traffic control and observability. For actuation mechanisms, SM supports functionalities such as traffic routing, traffic mirroring, traffic splitting, load balancing, rate limiting, circuit breaking, retries and inter-service response caching (see Figure~\ref{fig:SMFunct}). For observability mechanisms, SM offers functionalities such as traffic performance monitoring, distributed request tracing, endpoint health checks and outlier detection (see Figure~\ref{fig:SMFunct}).

As condition/options of control, SM supports timeouts and failover/fallback options for circuit breaking. Traffic policies can be specified and enforced in a hop-by-hop basis, either by considering request origin information or destination information (see Figure~\ref{fig:SMFunct}). For origin-oriented conditions, Istio for example supports options such as origin service labels, origin namespace, session information and request parameters or headers. For destination-oriented conditions, Istio supports options such as destination service labels, destination namespace, request endpoint and request port.

\subsection{Design Drivers and Non-functional Characteristics} 

SM enables the reduction of duplicated code implementing common logic for traffic management based on the sidecar proxy pattern. The sidecar proxy pattern allows traffic management to be application agnostic, i.e., agnostic to the language in which the application is developed and, to some extent, the logic of the application itself. In contrast, a proxy-less approach for SM is proposed in \cite{appswitchid}. For cloud-native applications built on the microservice-based architecture, a SM provides a way to compose a large number of discrete services into a centrally managed application. SM constitutes a highly manageable infrastructure layer enabling service-to-service communication, a relevant aspect of network softwarization.

One of the main design drivers of well-adopted SM implementations is related to DevOps practices. SM allows the decoupling of concerns between the application developer and the application operator. In this way, the configuration and maintenance of the service-to-service communication is not tied to the application code itself. Furthermore, SM supports more advanced functionality for Continuous Integration - Continuous Delivery/Deployment (CI/CD) such as A/B testing, blue/green and canary deployments.

In terms of traffic control and observability, SM supports flexible configurability thanks to the fine granularity levels that are supported in the definition of policy-based control conditions and actions. Policies are defined by following the so-called match-action abstraction, in which once a condition of control is met, a pre-defined action is matched and applied. More than just a cloud-native architectural pattern for traffic management, SM provides platform-level abstractions for easing and unifying traffic management of microservice-based applications.

Moving into more architectural matters, SM management, control and enforcement logics are decoupled into a Management Plane (MP), a Control Plane (CP) and a Data Plane (DP), respectively (see Figure~\ref{fig:general-architecture-service-mesh}). The MP, a.k.a. operational plane, is in charge of managing multiple SMs in a federated way. The MP can provide functionalities such as unified configuration and observability; discovery, interoperability and federation of multiple and heterogeneous SMs; and lifecycle management for SM and its applications, all under a single API.

The CP manages and configures the different components of the DP. It allows validation, ingestion, processing and distribution of configuration policies. It also monitors policy updates and propagates them to the DP components during runtime. The DP is usually composed by a set of lightweight proxies that are automatically injected and deployed along microservice instances, i.e., the sidecar proxy pattern. In more distributed deployment models, ingress, egress and/or east-west gateways are also part of the DP. DP components mediate communication on behalf of microservices and perform enforcement of traffic policies. 

Actuation logic is generally applied in an on/off basis. Recent research proposals study softer/smoother actuation mechanisms \cite{10.1145/3447851.3458740}. In terms of communication protocol support, SM implementations have limited support for Layer 4 / Layer 7 (L4/L7) communication protocols, i.e., L4 TCP, and L7 HTTP and gRPC.

As for network virtualization and abstraction matters, SM rely on the concept of network overlays such as the Container Network Interface (CNI). Furthermore, due to the sidecar proxy pattern, the traffic control and observability logic is run as a coupled process in user-space.

In terms of extensibility, Envoy-based DP components support WebAssembly (WASM) based extensions allowing flexible programmability of the proxy logic. On the other hand, many of the observability features of SM are based on external addons that can easily be inserted or removed. 

\begin{figure*}[htbp]
	\centering
	\includegraphics[width=1\linewidth]{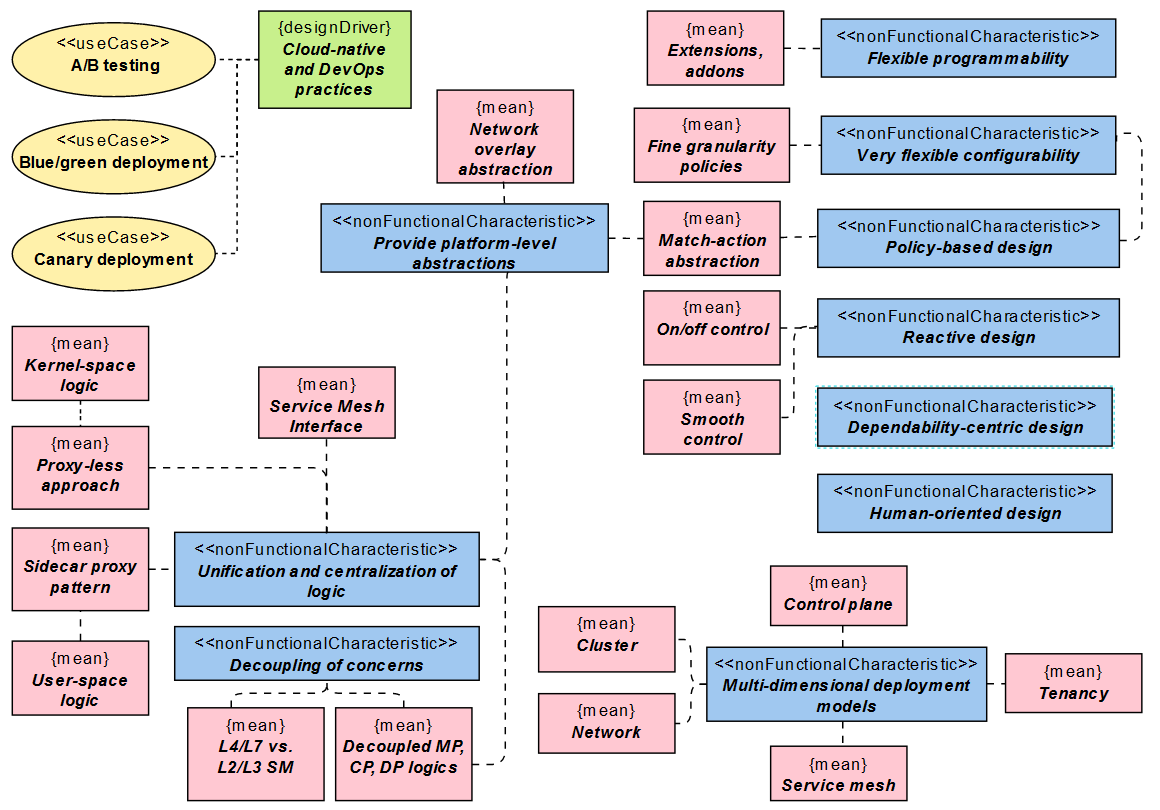}
	\caption{Design drivers and non-functional characteristics of SoTA Service Mesh}
	\label{fig:SMNonFunct}
\end{figure*}

In terms of deployment flexibility, Istio is one of the more mature implementations since it considers different dimensions related to various deployment models\footnote{https://istio.io/latest/docs/ops/deployment/deployment-models/, Istio: Deployment Models, accessed 2021-10-20}: i. the cluster, ii. the network, iii. the CP, iv. the SM and v. the tenancy dimensions. Figure~\ref{fig:SMNonFunct} summarizes SM's design drivers and non-functional characteristics.

\section{Qualitative Evaluation of Service Mesh-based Traffic Management}
\label{conceptualEvaluation}

This section discusses the main results of the systematic and qualitative analysis we performed by comparing the evaluation criteria from MEC and its workloads presented in Section~\ref{evaluationCriteria} vs. the characteristics and design drivers of SoTA SM presented in Section~\ref{SM}. This section also highlights limitations, tradeoffs and future research directions for the MEC-specialized SM we envision.

\subsection{Efficient Service Mesh Architecture}

Performance-demanding application workloads, in particular latency-sensitive applications, require the traffic management logic to be designed with minimum latency overhead. Similarly, resource-constrained individual edge clusters require minimum resource consumption. There are several aspects from the non-functional characteristics of SM that represent evident performance and resource costs.

\paragraph{Service mesh interface} As we mentioned before, the SMI initiative aims at providing a minimum set of core features targeting today's SM use-cases in the context of generic IT application workloads running in more CC environments. In this sense, this initiative aims at providing a standard interface towards external cloud-native technologies, e.g., Kubernetes, to interact with different implementations of SM and to support federation across SM instances. 

Nevertheless, current SM design lacks holistic or system-wide considerations related to the different domains and dimensions part of MEC; in particular, the design and standardization of a proper interface towards the MN to adequately implement functionalities such as end-to-end seamless mobility support or MN-aligned QoS assurance. To the best of our knowledge, we are the first ones to highlight this SM challenge.

\paragraph{Communication patterns and protocols} SM is mainly oriented towards the fourth generation of application connectors \cite{10.1145/3474624.3477072}, in which application components mainly communicate via RESTful APIs. The vast majority of generic IT application workloads today follow such communication pattern. However, this aspect represents a limitation in terms of supported communication patterns, which is also recognized by Kosińska, J. et al. \cite{kz20}. Emerging application workloads may require more performant communication patterns such as the delayered approaches that work at the system call level.

SM supports connection-oriented and reliable protocols such as HTTP and TCP which are more appropriate for generic IT applications. This is due to the fact that traditional IT application workloads are mainly characterized by their reliability requirements. However, such protocols may not be adequate for latency-sensitive applications since they may suffer delay-causing issues such as Head-of-Line (HOL) blocking. Protocols with lower transport delays such as QUIC, UDP and RTP are preferred by latency-sensitive applications. Antichi, G., et al. also acknowledge this limitation \cite{10.1145/3373360.3380834}.

\paragraph{Decoupled/split logic} There are intrinsic mechanisms part of the SM design intended to provide abstractions from the underlying system complexities. One of them is the decoupling of the traffic management into a MP, a CP and a DP, which is in part driven by the need of separation of concerns. Another related aspect is the fragmentation of SMs into L4/L7 class and L2/L3 class. On the other hand, SM relies on cloud-native network overlays. Such divide-and-conquer approaches have implications in terms of variable performance overheads that may not be appropriate for performance-demanding workloads; this issue is also identified by Antichi, G, et al. \cite{10.1145/3373360.3380834}.

\paragraph{Sidecar proxy pattern} The sidecar proxy pattern is widely adopted by SM to provide unified control and observability logic with very fine granularity. However, this approach has been criticized due to the implicated latency overhead that is present especially for traffic exchange across co-located microservice instances. The sidecar proxy pattern also brings performance overheads due to context switching between user-space and kernel-space. Antichi, G., et al. also recognize this challenge \cite{10.1145/3373360.3380834}. The added delays are exacerbated for longer and more complex microservice chains characteristic of MEC workloads.  

The sidecar proxy pattern also incurs resource consumption overhead due to the fact that the traffic management logic is replicated at a per- microservice instance level. Such extra resource consumption may not be appropriate for edge clusters due to their resource constraints. As pointed out by Dobaj, J., et al. \cite{10.1145/3361149.3361174}, both latency and resource overheads represent inherent scalability concerns associated with the SM architecture.

Furthermore, the fact that proxies are injected per- microservice instance creates dependencies between the sidecar proxies and the application in terms of life cycle management, which may not be appropriate for highly dynamic management of traffic and SM infrastructure due to its limited flexibility. 

On the other hand, since proxies implement both the actuation and instrumentation logic, this may represent a non-desired coupling between traffic control and observability, especially for automated traffic control, since such logics may have bi-directional performance implications.

The level of granularity required for traffic management of performance-demanding applications may not need to be the same than the one for generic IT applications. Especially, due to a tradeoff between the very fine granularity supported by SM for traffic control and observability vs. the incurred latency overhead of the sidecar proxies allowing such granularity levels. 

In the case of Istio version 1.11.4, every couple of proxies adds about 2.65 ms to the 90th percentile latency. In the DP, an Envoy proxy uses 0.35 vCPU and 40 MB memory for 1000 req/s. In the CP, Istiod (Istio's daemon consolidating control plane functionality into a single binary) uses 1 vCPU and 1.5 GB of memory\footnote{https://istio.io/latest/docs/ops/deployment/performance-and-scalability/, Istio Performance and Scalability, accessed 2021-11-28}.

Linkerd's own implementation of proxies, a Rust-based micro-proxy, is designed by considering performance requirements, thus service-to-service communication takes less than 1/5th the maximum extra latency taken by Istio while consuming 1/9th the memory and 1/8th the CPU when compared to Istio's DP\footnote{https://linkerd.io/2021/05/27/linkerd-vs-istio-benchmarks/, Benchmarking Linkerd and Istio, accessed 2021-11-04}. 

AppSwitch has been used as an Istio plugin to test performance enhancements. Results indicate an enhancement of over 18 times in the 50th percentile latency compared to vanilla Istio\footnote{https://istio.io/latest/blog/2018/delayering-istio/, Delayering Istio with AppSwitch: Automatic application onboarding and latency optimizations using AppSwitch, accessed 2021-11-01}. A similar proxy-less approach is recently starting to get adopted by Cilium\footnote{https://isovalent.com/blog/post/2021-12-08-ebpf-servicemesh, How eBPF will solve Service Mesh - Goodbye Sidecars, accessed 2021-12-10} based on eBPF, a kernel technology supporting custom programs to be run in kernel-space. However, the aforementioned performance metrics for Istio, Linkerd and AppSwitch are not comparable since they are not performed under the same conditions nor consider conditions characteristic of MEC setups. Nonetheless, performance enhancements from the kernel-space approaches need to be carefully evaluated in large-scale and highly distributed environments.

Table~\ref{tabArchSM} summarizes all of these challenges. To the best of our knowledge, we are the first ones to highlight the challenges in the table that do not have associated any reference.

\begin{table}[htbp]
\caption{Limitations associated with efficient Service Mesh architecture\label{tabArchSM}}
\centering
\begin{tabular}{|m{3cm} | m{8cm}|}
\hline
\textbf{Functional/ Non-functional Characteristic} & \textbf{Limitations} \\ [0.5ex] 
\hline
Service mesh interface & Lack of holistic interface design for end-to-end integration between the MN and the EC domains
\\
\hline
\multirow{2}{*}{\makecell[l]{Communication\\patterns and protocols}} & Limited communication pattern support; mainly for the 4\textsuperscript{th} generation of application connectors \cite{kz20}
\\
\cline{2-2}
& Lack of support for faster/telco-grade communication protocols and standards \cite{10.1145/3373360.3380834}
\\
\hline
\multirow{3}{*}{Decoupled/split logic} & Decoupled MP, CP and DP logic may introduce performance overheads due to the need of network-based communication across planes 
\\ 
\cline{2-2}
& L2/L3 SM and L4/L7 SM represent a split approach of the communication stack which may introduce performance overheads \cite{10.1145/3373360.3380834}
\\
\cline{2-2}
& Network overlay may introduce performance issues with variable delay overheads \\
\hline
\multirow{6}{*}{Sidecar proxy pattern} & Limited scalability and resource overhead due to replicated proxy logic per- microservice instance \cite{10.1145/3361149.3361174}
\\
\cline{2-2}
& Per-hop latency overhead even for communication across co-located microservice instances \cite{10.1145/3373360.3380834}
\\ 
\cline{2-2}
& Context-switching overhead since sidecar proxy logic is run in user-space \cite{10.1145/3373360.3380834}
\\
\cline{2-2}	
& Latency and resource overhead due to inadequate granularity level in control and observability logics \\
\cline{2-2}	
& Dependent and inflexible life-cycle management of proxies and microservice instances \\
\cline{2-2}
& Performance coupling between observability and actuation logics since both are implemented in a common sidecar proxy \\
\cline{2-2}
& Lack of quantification of performance costs associated with sidecar vs. sidecar-less approaches on large-scale applications hosted on edge cloud \\ [1ex]		
\hline
\end{tabular}
\end{table}

\subsection{Differentiated QoS Assurance Support in Service Mesh}
Performance demands of emerging applications require more efficient traffic steering mechanisms with stronger QoS assurance support. The fact that SM is not designed for deep EC is reflected in the lack of awareness, alignment, coordination and integration with the QoS assurance mechanisms proposed for the MN domain. Antichi, G., et al. also identify a general lack of support for QoS assurance \cite{10.1145/3373360.3380834}.

\paragraph{Dependability-centric traffic control} In terms of traffic control, most of the condition-based traffic actuation mechanisms supported by SM (see Figure~\ref{fig:SMFunct}) are intended to address reliability requirements of generic IT applications. This represents a good initial set of features for applications with strong dependability needs such as mission-critical applications. However, the traffic management needs from bandwidth-demanding applications and, most of all, the ones from latency-sensitive applications are not considered by such traffic control mechanisms. To the best of our knowledge, we are the first ones to highlight this SM challenge. 

There is a lack of holistic and differentiated QoS assurance mechanisms for the different types of application requirements of MEC workloads in alignment with the QoS assurance mechanisms of the MN and with awareness of the dynamic traffic performance through the MN path. Ways to address such limitations are end-to-end SFC for MEC application workloads similar to the ones proposed in \cite{9149045}\cite{9059340}, but rather based on L4/L7 SM. Other more appropriate actuation mechanisms may include prioritization of traffic and support for dynamic traffic shaping.

\paragraph{General-purpose multi-tenant isolation} SM considers both hard- and soft- multi-tenant isolation requirements of generic IT workloads in its deployment models and in its condition-based traffic control logic. However, there is no thorough consideration regarding isolation of the delivered performance to avoid performance implications across workloads with diverse characteristics and requirements. We consider that support for performance-oriented multi-tenant isolation similar to the concepts of network slicing would be needed in the MEC context. To the best of our knowledge, we are the first ones to highlight this SM challenge.

\paragraph{Zero-downtime rolling updates} Today's SM provides a set of actuation mechanisms such as traffic mirroring, traffic splitting and traffic redirection (see Figure~\ref{fig:SMFunct}) intended to be used for zero-downtime rolling updates across application versions\footnote{https://blog.sebastian-daschner.com/entries/zero-downtime-updates-istio, Zero-Downtime Rolling Updates With Istio, accessed 2021-09-01}. We envision that such actuation mechanisms can be exploited with a different purpose in MEC workloads. As mentioned before, one of the main challenges of MEC is the highly dynamic traffic conditions such as the ones generated upon mobility events. Actuators originally designed for providing zero-downtime application rollouts can be leveraged to provide seamless mobility support with minimum QoE degradation. To the best of our knowledge, we are the first ones to highlight this SM opportunity.

\paragraph{Human-oriented traffic observability} In terms of traffic observability, the instrumentation and data collection mechanisms implemented in the SM are mainly intended to be used by human system operators. Nevertheless, such proposed mechanisms may not necessary be appropriate for automated traffic management. To the best of our knowledge, we are the first ones to highlight the following SM challenges.

The per-hop, per-endpoint granularity associated with traffic performance metrics may not be appropriate for differentiated QoS assurance since it does not consider aggregated performance estimations at the level of end-to-end service chains.

Furthermore, the frequency of performance metric collection may not be fast enough to appropriately detect and react upon fast changing traffic load conditions. In the case of Prometheus, for example, it is recommended not to have a scraping interval lower than 15s due to Kubelet's resource usage metric resolution. Envoy's traffic performance metrics can be flushed into stats sinks with a minimum interval of 1ms; however, it is unknown if Prometheus would scale well upon high traffic load and large-scale deployments.

To perform automated traffic control, other system metrics need to be considered from different parts of the application and the underlying execution environment \cite{kz20}. On the other hand, the way in which traffic observability is provided via external addons, as in Istio, offers poor configurability of the observability logic. 

Table~\ref{tabQoSSM} summarizes all of these challenges.  To the best of our knowledge, we are the first ones to highlight the challenges in the table that do not have associated any reference. 

\begin{table}[htbp]
\caption{Limitations associated with QoS assurance support in Service Mesh\label{tabQoSSM}}
\centering
\begin{tabular}{|m{3cm} | m{8cm}|}
\hline
\textbf{Functional/ Non-functional Characteristic} & \textbf{Limitations} \\ [0.5ex] 
\hline
\multirow{6}{*}{\makecell[l]{Dependability-centric\\traffic control}} & Limited dependability-centric design with focus on reliability requirements \\
\cline{2-2}
& Unawareness of diverse application performance requirements \\
\cline{2-2}
& Lack of differentiated traffic routing logic based on application requirements \\
\cline{2-2}
& Lack of actuation, conditions of control and observability mechanisms for other purposes than merely reliability support \\ 
\cline{2-2}
& Lack of traffic shaping mechanisms for bandwidth-demanding applications \\
\cline{2-2}
& Lack of traffic prioritization mechanisms for latency-sensitive applications \\
\hline
General-purpose multi-tenant isolation & No support for performance-oriented, multi-tenant isolation \\
\hline
Zero-downtime rolling updates & Lack of seamless mobility support with minimum QoE degradation \\
\hline
\multirow{6}{*}{\makecell[l]{Human-oriented\\traffic observability}} & Intended to be for human operator's visualization \\
\cline{2-2}
& Inappropriate level of granularity in performance metrics, not at the service chain level required for QoS assurance \\ 
\cline{2-2}
& Lack of end-to-end performance estimation \\
\cline{2-2}
& Limited frequency of metric collection, not appropriate for very dynamic changes in metrics values \\
\cline{2-2}
& Inflexible and cumbersome configurability of metric collection due to poor integration with external addons \\
\cline{2-2}
& Performance metrics are limited to traffic monitoring and tracing; they need to be complemented with system performance metrics \cite{kz20}
\\ [1ex]		
\hline
\end{tabular}
\end{table}

\subsection{Autonomous Service Mesh}
Intelligence, automation and autonomy are desired characteristics of cloud-native traffic management approaches to cope with the strong traffic performance guarantees MEC workloads require. There is a need for having more abstracted, automated and autonomous mechanisms to efficiently adapt not only the ongoing traffic but also the SM architecture itself to the highly dynamic conditions characteristic of emerging applications and their underlying execution environment.

\paragraph{Flexible configurability} Today's SM offers very fine granularity in terms of traffic control policies and it also supports many dimensions of configurability in their deployment models to provide the very flexible configurability required by generic IT use-cases. Nevertheless, SM frameworks are still considered complex to operate; such levels of flexibility tamper their adoption due to the inherent degrees of complexity, as mentioned by Dobaj, J., et al. \cite{10.1145/3361149.3361174}.

Despite all efforts to make traffic management easier, we consider that current levels of abstractions are just the first step towards higher levels of autonomy in SM. Automation and autonomy could offer higher levels of abstraction with more human-intuitive mechanisms than the ones provided by policy-based traffic management, for example.

\paragraph{Complexity abstractions}

The match-action abstraction has historically been used by human network operators. Such traffic control mechanism is more suitable for rule-based systems dealing with relatively less dynamic and simpler traffic conditions. However, this mechanism may not be appropriate to build more advanced, automated and autonomous traffic management approaches which need to address the strong performance demands of future workloads under highly dynamic conditions.

Furthermore, different approaches used for complexity abstractions such as network overlays and the sidecar proxy pattern imply certain performance costs. There is a need to quantify the associated performance cost vs. performance gains of automated/autonomous traffic management mechanisms. To the best of our knowledge, we are the first ones to highlight the mentioned SM challenges.

\paragraph{Decoupled policy propagation} The SM CP takes care of propagating policy updates to the different components of the DP to deal with changes in traffic control stipulated by human operators. However, such decoupled mechanism may not be able to support frequent policy reconfigurations associated with automated/autonomous traffic management in scalable and performant ways. Highly dynamic reconfigurations may lead to network bottlenecks when the CP propagates frequent updates across DP components. To the best of our knowledge, we are the first ones to highlight this SM challenge.

\paragraph{Human-oriented management plane} Today's SM MP only takes care of SM operational matters with very poor levels of automation and autonomy. As recognized by Antichi, G., et al. \cite{10.1145/3373360.3380834}, we also envision a SM capable of supporting architectural adaptability and on-the-fly configuration of communication protocols and patterns in response to highly dynamic and diverse requirements of MEC workloads.

\paragraph{Reactive traffic control} Traffic control mechanisms supported by today's SM implementations are reactive; the actuation logic is triggered upon the detection of a set of conditions in an on/off fashion. There is a lack of proactive and smooth actuation mechanisms which may be enabled by the introduction of, for example, analytics and inference layers \cite{8705911}. Mendon\c{c}a, N., et al. also state that none of the current SM solutions support general-purpose self-adaptation capabilities \cite{10.1145/3474624.3477072}. Saleh Sedghpour, M. R., et al. also identify the lack of adaptable traffic control mechanisms such as smooth circuit breaking actuation logic \cite{10.1145/3447851.3458740}.

Table~\ref{tabAutoSM} summarizes all of these challenges. To the best of our knowledge, we are the first ones to highlight the challenges in the table that do not have associated any reference.

\begin{table}
\caption{Limitations associated with autonomous Service Mesh\label{tabAutoSM}}
\centering
\begin{tabular}{|m{3cm} | m{8cm}|}
\hline
\textbf{Functional/ Non-functional Characteristic} & \textbf{Limitations} \\ [0.5ex]
\hline
\multirow{2}{*}{\makecell[l]{Flexible\\configurability}} & Complex to operate which tamper their adoption \cite{10.1145/3361149.3361174} \\
\cline{2-2}
& Fine granularity level in configuration imply performance overheads \\
\cline{2-2}
& Lack of higher levels of abstraction e.g., intent-driven configuration \\
\hline
\multirow{2}{*}{\makecell[l]{Complexity\\abstractions}} & Match-action abstraction limited to rule-based systems and not appropriate for autonomous management \\
\cline{2-2}
& Complexity abstractions imply performance costs; there is a need to evaluate performance cost vs. performance gain of higher abstraction levels \\
\hline
\makecell[l]{Decoupled\\policy propagation} & Not designed to support frequent policy reconfigurations in scalable and performant ways \\
\hline
\multirow{2}{*}{\makecell[l]{Human-oriented\\management plane}} & SM management plane limited to human-oriented operational matters \\
\cline{2-2}
& Lack of self-adaptable functionality and architecture \cite{10.1145/3474624.3477072}\cite{10.1145/3373360.3380834}
\\
\hline
\multirow{2}{*}{\makecell[l]{Reactive\\traffic control}} & Actuation and condition mechanism limited to on/off, lack of softer/smoother control mechanisms \cite{10.1145/3474624.3477072}\cite{10.1145/3447851.3458740}
\\
\cline{2-2}
& Lack of analytics and inference layers with more predictive and proactive logics \cite{8705911}
\\ [1ex]		
\hline
\end{tabular}
\end{table}

\bibliographystyle{unsrtnat}
\bibliography{references}  



\end{document}